# Dirac cone spectroscopy of strongly correlated phases in twisted trilayer graphene

Cheng Shen[1,*], Patrick J. Ledwith[2], Kenji Watanabe[3], Takashi Taniguchi[3], Eslam Khalaf[2], Ashvin Vishwanath[2] and Dmitri K. Efetov[1,*]

1. ICFO - Institut de Ciencies Fotoniques, The Barcelona Institute of Science and Technology, Castelldefels, Barcelona, 08860, Spain

2. Department of Physics, Harvard University, Cambridge, MA 02138, USA

3. National Institute for Materials Science, 1-1 Namiki, Tsukuba, 305-0044, Japan

*Correspondence to: cheng.shen@icfo.eu; dmitri.efetov@icfo.eu

**Mirror-symmetric magic-angle twisted trilayer graphene (MATTG) hosts flat electronic bands close to zero energy, and has been recently shown to exhibit abundant correlated quantum phases with flexible electrical tunability.**[1–3] **However studying these phases proved challenging as these are obscured by intertwined Dirac bands. In this work, we demonstrate a novel spectroscopy technique, that allows to quantify the energy gaps and Chern numbers of the correlated states in MATTG by driving band crossings between Dirac cone Landau levels and the energy gaps in the flat bands. We uncover hard correlated gaps with Chern numbers of $C = 0$ at integer moiré unit cell fillings of $\nu = 2$ and $3$ and reveal novel charge density wave states originating from van Hove singularities at fractional fillings of $\nu = 5/3$ and $11/3$. In addition, we demonstrate the existence of displacement field driven first-order phase transitions at charge neutrality and half fillings of the moiré unit cell $\nu = \pm 2$, which is consistent with a theoretical strong-coupling analysis, implying the breaking of the $C_2T$ symmetry. Overall these properties establish the diverse and electrically tunable phase diagram of MATTG and provide an avenue for investigating electronic quantum phases and strong correlations in multiple-band moiré systems.**

Fundamental signatures of novel quantum phases often emerge as gap openings, which signify phase transitions in band topology, electronic correlations and elementary excitations. Here flat band moiré systems have recently emerged as a robust and tunable platform to investigate strongly correlated electrons, where exotic gapped quantum states of correlated insulators,[4–12] superconductivity,[13–15] and topological Chern insulators[12,16–24] were uncovered. In particular the recently realized mirror-symmetric twisted trilayer graphene system with a magic angle $\theta \sim 1.5°$ (MATTG) has evolved as an interesting new platform (Fig. 1a right). Its energy spectrum however contains a coexisting moiré flat band and a steep Dirac cone (Fig. 1b left), which makes the detection of correlation-induced insulating states through transport very difficult. Explicitly detecting and quantifying these nontrivial gapped states in the flat bands of MATTG requires a novel technique that can untie the electrons in the intertwined bands.

In this work, we overcome these challenges by introducing a novel Dirac cone spectroscopy approach, which allows to quantify the interaction drive gaps in the flat bands, by studying the band crossings of these with the Landau levels of the Dirac cone under small perpendicular magnetic field $B$. Fig.1a shows the device schematics of the double gated ultra-clean MATTG device with a twist-angle of $\theta \approx 1.5°$ (see SI), where the two gates allow to induce a displacement field $D$ and also change the electron filling per moiré unit cell $\nu$. Fig.1c shows the measured phase diagram of longitudinal resistance $R_{xx}$ vs. $\nu$ and $D$ at a temperature of $T = 35$ mK, which shows a multitude of correlated states at various integer fillings and two robust superconducting regions in the interval of $|\nu| = 2 + \delta (0 < \delta < 1)$, in good analogy to previous work[1,2].

Figs.1d and e show the color plots of $R_{xx}$ and the Hall resistance $R_{xy}$ vs. perpendicular magnetic field $B$ and $\nu$ in the conduction moiré band, and Figs.1f shows the corresponding line-cuts of $R_{xy}$. In addition to the typically observed insulators and Landau levels (LL) in the flat-bands, which originate at different integer fillings, we also observe an extra set of LLs which are associated with the Dirac band[1,2] (D-LLs), and emerge from charge neutrality at much lower magnetic fields. These D-LLs cross the states of the flat band in multiple positions in the $\nu$-$B$ phase-space and form a complex crossing pattern, which is defined by the overlap between the correlated gaps and LLs in the flat bands and the gaps between the different D-LLs. Fig. 1b (right) exemplifies the main features at a filling of $\nu = 2$ in Figs. 1d and e, by showing the corresponding band schematics depending on the exact values of $\nu$ and $B$.

Here the crossing points in a non-hybridized situation follow the simple rule that $\mu^f = \mu^D$ and the total Chern number $C = C^f + C^D$, where $\mu^f$ and $\mu^D$, $C^f$ and $C^D$, are chemical potential and Chern number for the flat band and Dirac cone, respectively. In a perpendicular magnetic field $B$, the Dirac cone is replaced by the discrete D-LLs with Chern numbers $C^D = 4N + 2$, which have a chemical potential $\mu_N^{D-LL} = \mu_{N=0}^{D-LL} + v_F\sqrt{2e\hbar N sgn(N)B}$ (here, $e$, $\hbar$, $N = 0, \pm 1, \pm 2$ ... and $v_F$ are elementary charge, Planck constant, Landau level index and Fermi velocity of Dirac cone, respectively)[25,26]. These considerations allow us not only to directly extract the Chern numbers of the correlation induced gaps in the flat bands via the relation $C^f = C - (4N + 2)$, but also to quantify these. By tracking the zeroth D-LL ($N = 0$) in the $\nu$-$B$ phase diagram, the chemical potential of the flat band $\mu^f$ relative to the Dirac cone vertex can be extracted as $\mu^f - \mu_{N=0}^{D-LL}$[1], which shows a large energy scale and linear background with respect to charge filling $\nu$, as a consequence of shifting of the Dirac cone with charge filling[27]. Pinning of the chemical potential is a thermodynamic signature of flavor symmetry breaking[23,28], which helps to determine the exact integer fillings of the flat band.

### Extraction of Chern numbers

Interestingly, we find that at most integer fillings, in particular at $\nu = 2, 3$ and at fractional filling $\nu = 5/3$ the Hall resistance $R_{xy} = h/Ce^2$ shows well quantized plateaus (Fig. 1f) which follow linear trajectories $dn/dB = Ce/h$ in the $\nu$-$B$ phase diagram, in agreement with the Streda formula (Fig. 1 e and d and SI). These plateaus correspond to Chern numbers $C = 2, 6, 10$, depending on the magnitude of $B$ (phase III in Extended Data Fig. 1), which is precisely what is expected for hard interaction driven gaps in the flat bands with a Chern number of $C^f = 0$, which are superimposed by D-LLs. The exact quantization of $R_{xy}$ indicates transport through the edge states formed between the D-LL, while the electrons in the flat bands are fully localized. The corresponding $R_{xx} = 0\Omega$ regions are flanked by two peaks that are generated by the scattering between D-LL edge states and unlocalized flat band charges, and so mark the edges of the gapped flat bands. As we will further discuss in much greater detail, we identify the $\nu = 2, 3$ states as $C^f = 0$ correlated insulators, while the $\nu = 5/3$ is likely a charge density wave state. Analysis for other fillings also shows a metallic state at $\nu = -3$, a tiny or nodal gap at $\nu = \pm 1, -2$ and electron solid state at $\nu = 11/3$ (Extended Data Fig. 2, Extended data Fig. 3 and see SI section B).

The observed $C^f = 0$ at $\nu = 2, 3$ emerges at a quite low magnetic field $B < 0.3T$ implying a $C^f = 0$ zero-field ground state of the flat bands, and survives to a large perpendicular magnetic field $B > 8T$ without transition to $|C^f| = 4 - |\nu|$ which is formed by sequentially filling eight valley-projected flat Chern bands with time-reversal symmetry breaking (Extended Data Fig. 2)[19–24]. For the strong ground state of $C^f = 0$, one candidate mechanism is translation symmetry breaking which doubles the number of bands and leads to a natural description with eight $C^f = 0$ bands and eight

$C^f = \pm 1$ bands; there are then natural states at $\nu = 3$ with zero total $C^f$ that are obtained by filling combinations of these sixteen bands.[29] Alternatively, $C^f = 0$ states at both non zero fillings $\nu = 2, 3$ are also consistent with the recently proposed time-reversal symmetric "incommensurate Kekulé spiral" (IKS) order which appears in the presence of strain.[30] Further studies are needed to uncover the competitions of these Chern insulator states in MATTG which are determined by symmetry breaking that is sensitive to specific parameters like strain and magnetic field.

**Extraction of gap values and their $D$-dependence**

We further estimate the sizes of the interaction induced gaps. Such spectroscopy can be achieved by measuring the chemical potential jump across these gaps, which can be achieved by measuring the crossing points in $\nu$-$B$ of its band edges with two consecutive D-LLs, as is demonstrated in Fig. 2a for the $\nu = 2$ state. Fig. 2b shows the schematic of the band crossings between the D-LLs and the flat band gap. To obtain the exact band edges $A_N$ and $B_N$ with the $N$th D-LL cross, we track the crossing state that corresponds to partial filling of D-LLs at Fermi level, which is marked by a $R_{xx}$ peak. When this state occurs between the D-LLs and the correlated gap (phase I in Extended Data Fig. 1), it features a transition between two adjacent $R_{xy}$ plateaus with $C = 4N \pm 2$ under changing magnetic field $B$. When this phase reaches the crossing between the D-LL and the flat band edges, the $R_{xy}$ plateau transition disappears and the $R_{xx}$ peak merges into the two flanking peaks that manifest edges of the flat band. The so extracted magnetic fields at the band edges $B_{A_N}$ and $B_{B_N}$ determine the corresponding chemical potentials $\mu_A$ and $\mu_B$ relative to zeroth D-LL, and so to estimate the gap size of the correlated states via $\Delta = \mu_A - \mu_B = v_F\sqrt{2e\hbar N sgn(N)B_{A_N}} - v_F\sqrt{2e\hbar N sgn(N)B_{B_N}}$, where for $\nu = 2$ we find a $\Delta(\nu = 2) \approx 4.8 meV$ for $D = 0$.

We repeat this spectroscopy for different values of $|D| < 0.2$ V/nm, where $v_F$ is not significantly altered (see SI section C), and find that $\Delta(\nu = 2)$ shows enhancement with respect to increased $D$ (Fig. 2c and Extended Data Fig. 5). Strong-coupling analytics and Hartree-Fock simulations can explain this trend: for small displacement fields and not-too-large twist angles, the dispersion of the twisted bilayer graphene (TBG) subsystem bands is increased. This increased dispersion can combine with the interaction-induced gap such that the total gap is increased (see SI section J for a detailed discussion). In previous reports on MATTG[1,2] and also our device, the nearby superconductor at $\nu = 2 + \delta$ shows $T_c$ enhancement in moderate displacement field as well. This observed increase in both $\Delta(\nu = 2)$ and $T_c$ with displacement field, hints at a relation between correlated insulator and nearby superconductor, and is compatible with a scenario where the pairing scale is controlled by the TBG subsystem dispersion.[37]

In Fig. 2c, notably, we also found that the chemical potential $\mu_B - \mu_{N=0}^{D-LL}$ at $\nu = 2$ decreases in increased displacement field. This should be interpreted as an upward shift in energy $\mu_{N=0}^{D-LL}$ of the Dirac cone vertex; the flat band energy $\mu_B$ is not expected to change much in a moderate displacement field.[38] Single particle simulations predict that the Dirac cone immediately hybridizes with the flat bands for arbitrarily small displacement fields as a result of the broken mirror symmetry lifting the degeneracy. The degeneracy lifting leads to two Dirac points that are equal-weight superpositions of the two mirror sectors; i.e. the flat bands and Dirac cone are maximally hybridized at these new Dirac points. The energies of these hybridized Dirac points move away from charge neutrality linearly with respect to $D$[33,34]. Our band crossing results at weak to moderate displacement field instead find a shifting energy $\Delta E = -\Delta(\mu_B - \mu_{N=0}^{D-LL})$ that is roughly quadratic in $D$ which suggests a hybridzation between the Dirac cone and flat band that is proportional to $D$ and not immediately maximal. This failure of the single particle picture is explained with our strong-coupling and Hartree-Fock predictions (SI section J).

The shifting of the Dirac cone induces a transition of the D-LL Chern number $C^D = 4N + 2$ in displacement field when Fermi level and magnetic field $B$ are fixed. In Fig. 3a, Hall resistance $R_{xy}$ plateaus at $\nu = 3$ and $\nu = 4$ show smaller total Chern number $C = C^f + C^D$ and thereby a lower total charge density in a higher $D$, substantiating that the Dirac cone also shifts upward in energy at $\nu = 3$ and $\nu = 4$ (Fig. 3c). We define the critical displacement field $D_c(\nu)$ to be the field where zeroth D-LL (Dirac cone vertex) reaches the Fermi level $\mu_B$ of integer fillings $\nu = 2, 3, 4$, i.e. $\mu_B - \mu_{N=0}^{D-LL} = 0$. At $D_c(\nu)$, there are $R_{xy}$ transitions from the $R_{xy}$ plateaus with $C = 2$ to divergent $R_{xy}$ at $\nu = 2, 4$ (Fig. 3a). At $\nu = 3$, the $R_{xx}$ dip that is produced by the zeroth D-LL edge state (Fig. 3b) disappears under changing displacement field through $D_c(\nu)$. According to such $R_{xx}$ and $R_{xy}$ transition, we acquire $D_c(\nu = 2) \approx D_c(\nu = 4) \approx 0.38 V/nm$ (Fig. 3a), $D_c(\nu = 3) \approx 0.48 V/nm$ (Fig. 3b). The acquired $D_c(\nu = 2)$ is well consistent with the chemical potential results in Fig. 2c where $\mu_B - \mu_{N=0}^{D-LL}$ at $\nu = 2$ is expected to reach zero at $D \approx 0.4 V/nm$ according to the quadratic fitting of $\Delta E \propto D^2$. The upward shifting of the Dirac cone changes the number of non-hybridized Dirac cone electrons and therefor impacts the overall transport behavior. At $D = D_c(\nu)$, the number of Dirac cone charges is at a minimum. As a consequence, resistance peak at zero magnetic field is expected to emerge at $D_c(\nu)$ for corresponding integer filling $\nu$, which is consistent with our experimental observation in Fig. 4a. Note that the reduction of Dirac electrons by Dirac cone shifting also reduces screening of the Coulomb interaction between flat band electrons: this effect would also lead to a correlated gap increase with displacement field.

What's the displacement field where Dirac cone is highly hybridized with flat band at specific integer filling? We found at $|D| > |D_c(\nu)|$ ($\nu = 2, 3, 4$), there is no signature of negative total Chern number $C$ that could appear from the hole side of the Dirac cone if Dirac cone were to keep shifting with increasing displacement field while remaining only perturbatively hybridized with the flat bands (Fig. 3a). In addition, $\nu = 4$ shows constant resistance of zero magnetic field at $|D| > |D_c(\nu = 4)|$ (Fig. 4a), contradictory to a non-hybridized Dirac cone shifting picture. These features point to that Dirac cone is strongly hybridized with flat band when Dirac cone vertex reaches near the Fermi level of the corresponding integer fillings $\nu = 2, 3, 4$. The hybridization at $\nu = 4$ forms stable Dirac cone point that is responsible for the constant resistance in higher displacement field. For $\nu = 0$, Dirac cone vertex is pinned at zero energy in displacement field by particle-hole symmetry (Extended Data Fig. 10 and SI section J), giving rise to the observation of constant resistance in a wide regime of moderate displacement field crossing $D = 0$ (Fig. 4a). Identifying the hybridization of Dirac cone via Chern number transition in displacement field at $\nu = 0$ is therefore not available in our experiment.

**Charge density waves and Fermi surface nesting**

In Fig.3a, pronounced Hall resistance sign change indicates emergent van Hove singularities (VHSs) on the hole side of $\nu = 2$ and $\nu = 4$ at high displacement field. At lower displacement field, VHSs shift to $\nu = 5/3$ and $\nu = 11/3$, and Hall resistance sign changes evolve to plateaus with $C = 4N + 2$ due to the coexisting Dirac cone. This strongly indicates a VHS origin for electron solid states at $\nu = 5/3$ and $\nu = 11/3$. VHS instability generates strong electronic correlation due to its high density of state (DOS) and in this case is expected to induce a charge density wave with moiré translational symmetry breaking. In our Hartree Fock simulation for the valence band of $\nu = 2$ (see more details in SI section J), we found VHS where six hole pockets merge into a single annular Fermi surface. $\nu = 5/3$ is extremely close to these VHS in energy and its annular Fermi surface is nearly nested with respect to wavevectors which connect moiré $K$ and $K'$ points. These wavevectors are the translation symmetry breaking wavevectors that correspond to a $\sqrt{3} \times \sqrt{3}$ tripling of the moiré unit cell in a $C_3$ symmetric way (Fig.3d). Breaking $C_3$ symmetry is believed to destroy CDW state by ruining the above nesting. This accounts for the rare observation of CDW in other devices as $C_3$ symmetry is often broken by strain effect. We found two neighboring regions in our device that show distinct behavior of longitudinal magnetoresistance $R_{xx}$ at $\nu = 5/3$: one features $R_{xx}$ peak indicating a few unfrozen flat band charges, i.e. CDW is incipient; while another features $R_{xx}$ dip, indicating CDW is well-developed (Fig. 1d and Extended Data Fig. 5). These distinct behaviors, together with expected sensitivity to $C_3$ symmetry, suggests an inhomogeneously distributed strain in our device.

The critical temperatures for correlated states at various fillings $\nu$ follow the hierarchy $T(\nu = 1\ \&\ 2\ \&\ 3) > T(\nu = 5/3\ \&\ 11/3)$ (Extended Data Fig. 6). In Fig. 3e, the melting of correlated states in increased temperature, manifested by vanishing of Hall resistance, shows a faster speed at $\nu = 5/3$ than $\nu = 2$, in accordance with the expected smaller inter-site Coulomb interaction responsible for fractional filling states as compared to stronger on-site Coulomb interaction for integer fillings.

### $D$-induced phase transitions

At $|D| > |D_c(\nu)|$, our spectroscopy is not available for the corresponding filling due to the strong hybridization of Dirac cone with the flat bands. On the other hand, the absence of a non-hybridized Dirac cone makes the gap of the whole system visible via temperature-dependent transport behavior. We found that both charge neutrality and $\nu = 2$ experiences a transition from metallic state to a thermal-activation insulating state in displacement field (Fig. 4a, Fig.4b, Fig. 4c). At $|D| \approx 0.77V/nm$, the insulating states are fitted with a thermal-activation gap $\Delta \approx 1.18 meV$ at charge neutrality and $\Delta \approx 0.26 meV$ at $\nu = 2$ (Fig. 4c). This means the hybridized Dirac cone must be gapped out by spontaneous breaking of $C_2T$ symmetry. We propose a first-order transition driven by a tunable competition between energetically competitive flat band symmetry breaking orders; in particular the Kramer intervalley coherence (KIVC), valley Hall (VH), valley polarization (VP) and semimetal (SM) states at charge neutrality and $\nu = 2$.[39–41] Fig. 4d plots the energies of these orders tuned by displacement field with numeric Hartree-Fock simulations, where KIVC is gapless and dominates at moderate displacement field and VH, which breaks $C_2T$, is fully gapped for $D > 0$ and is lower in energy for larger displacement fields before ultimately a symmetric semimetal becomes the ground state at the largest fields (Extended Data Fig. 10 and SI section J). The displacement field regime of gapped charge neutrality and $\nu = 2$ observed in our experiment is consistent with the theory prediction (Fig. 4a and SI section J). Notably, superconductivity is weakened at the regime of displacement field where VH is developed (Fig. 1c), consistent with the absence of superconductivity in samples with explicit $C_2T$ breaking with, say, an aligned hBN substrate.[19,42] This is also compatible with the theoretical proposals based on skyrmion pairing or pseudospin fluctuations.[37,43]

We note in a narrow regime of displacement field near $|D| = 0.34V/nm < |D_c(\nu = 4)|$, $\nu = 4$ shows insulating behavior at zero magnetic field as well (Fig. 4b). This is in contrast to what we expect that at such displacement field a small quantity of non-hybridized Dirac cone electrons should produce a metallic behavior. In perpendicular magnetic field, these Dirac cone electrons are distributed in D-LLs which composes of edge and localized states, and as a result, they are not interacted with localized flat band electrons when flat band is gapped, giving rise to $R_{xy}$ plateau with $C = 4N + 2$. However, without magnetic field, these Dirac cone electrons are itinerant and therefor may interact with localized flat band electrons via magnetic interactions such as Kondo scattering.[44] This is also possibly responsible for the gapped states at displacement field that is smaller than $D_c(\nu)$ for integer filling $\nu = 2$ reported previously.[45] Such candidate heavy fermion state in MATTG requires further studies on localized magnetic moments in flat band at these integer fillings.

In summary, we put forward a novel Dirac cone spectroscopy in MATTG that is employed to disclose electron solid states and quantify correlated gap with zero Chern number at integer or fractional fillings for flat band electrons. Such spectroscopy casts light on the exotic phase transition and complex correlated band structure subject to displacement field. Our work helps to bridge between flat band correlated insulator and nearby unconventional superconductor, and provides an avenue to quantitatively study flat band correlated insulators in other multiband moiré systems.[36,45,46]

**Data availability**

The data are available from the corresponding author upon reasonable request.

**Acknowledgement**

We thank B. Andrei Bernevig, Lede Xian and Quansheng Wu for fruitful discussions and Ipsita Das, Alexandre Jaoui, Chang-Woo Cho and Benjamin A. Piot for the help in cryogenic measurements. P. J. L would like to acknowledge fruitful discussions with Maine Christos and support by the Department of Defense (DoD) through the National Defense Science and Engineering Graduate Fellowship (NDSEG) Program. E. K. was supported by the German National Academy of Sciences Leopoldina through Grant No. LPDS 2018-02. A. V. was supported by a Simons Investigator award and by the Simons Collaboration on Ultra-Quantum Matter, which is a grant from the Simons Foundation (Grant No. 651440). D.K.E. acknowledges support from the Ministry of Economy and Competitiveness of Spain through the "Severo Ochoa" program for Centres of Excellence in R&D (SE5-0522), Fundació Privada Cellex, Fundació Privada Mir-Puig, the Generalitat de Catalunya through the CERCA program, funding from the European Research Council (ERC) under the European Union's Horizon 2020 research and innovation programme (grant agreement No. 852927). K.W. and T.T. acknowledge support from the Elemental Strategy Initiative conducted by the MEXT, Japan (Grant Number JPMXP0112101001) and JSPS KAKENHI (Grant Numbers 19H05790, 20H00354 and 21H05233).

**Author contribution**

C.S. and D.K.E. conceived of the project. C.S. fabricated devices, performed transport measurements and analyzed the experimental data. P.J.L, E.K. and A.V. performed the numeric simulations. K.W. and T.T. provided the hBN crystals. C.S., P.J.L., E.K., A.V. and D.K.E. discussed the data. C.S., P.J.L., E.K., A.V. and D.K.E. wrote the paper.

**Competing interests**

The authors declare no competing interests.

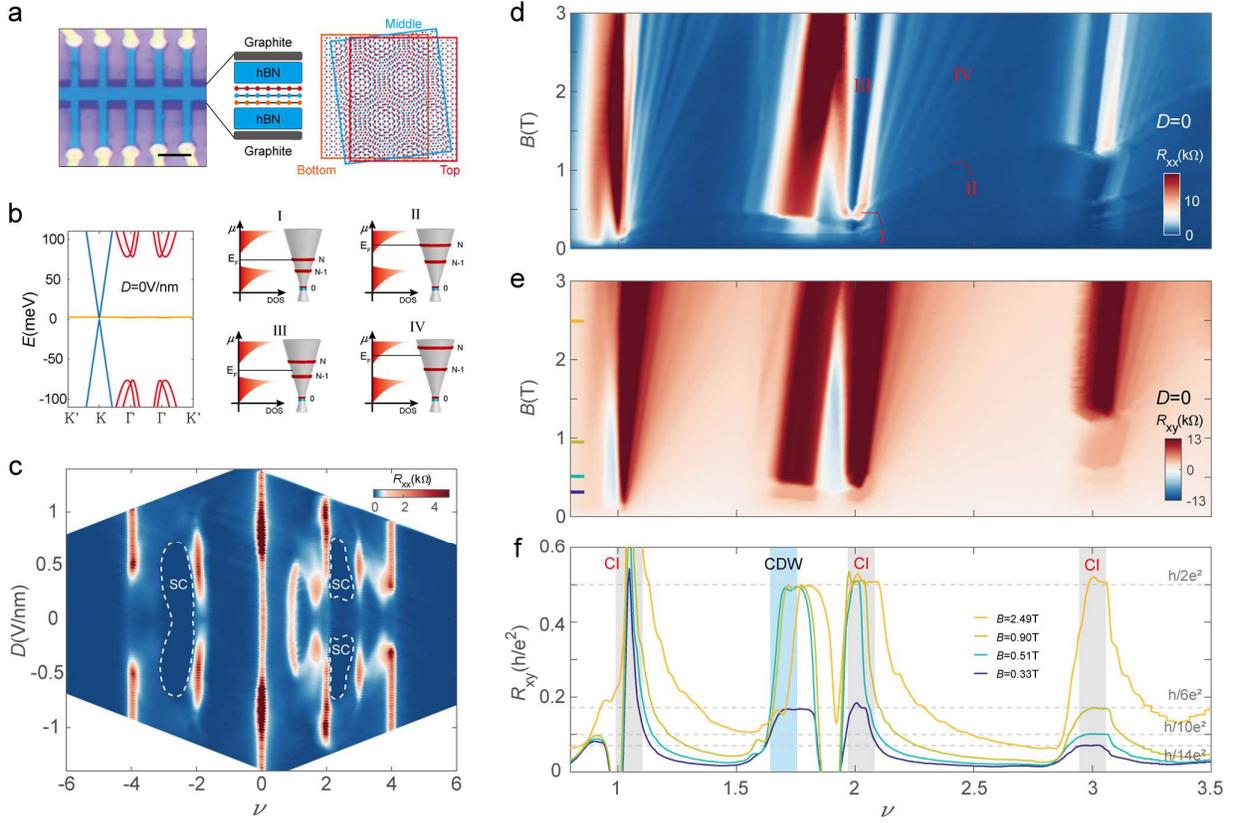

**Fig. 1 | Electrically tunable phase diagram, zero Chern numbers and electron solid states in flat band of magic angle twisted trilayer graphene (MATTG). a,** schematic of dual-gated MATTG devices. MATTG consists of three monolayer graphene which are sequentially stacked with twist angle 1.5° and −1.5° and are encapsulated by two hexagonal boron nitride flakes. Graphite flakes are used as top and bottom gates. The scale bar corresponds to 3μm. **b,** Left: single-particle band structure at $D = 0V/nm$ Bands colored blue, red and orange denote Dirac cone, moiré flat band and remote dispersive bands, respectively. Right: Interplay between flat band gaps and Dirac cone Landau levels, highlighting different scenarios in Fig. 1d. **c,** color plot of four-probe resistance as a function of displacement field $D$ and charge filling $\nu$ of moiré superlattice. Superconducting regions are surrounded by white dash curves. At the electron side, the superconductor onsets at nonzero displacement field. **d, e,** Landau fan diagram shown by longitudinal resistance $R_{xx}$ and transverse Hall resistance $R_{xy}$ at zero displacement field. **f,** linecuts of $R_{xy}$ are taken from (e). Gray and blue shading denotes correlated insulator (CI) and charge density wave (CDW) at integer fillings $\nu = 1,2,3$ and fractional filling $\nu = 5/3$, respectively.

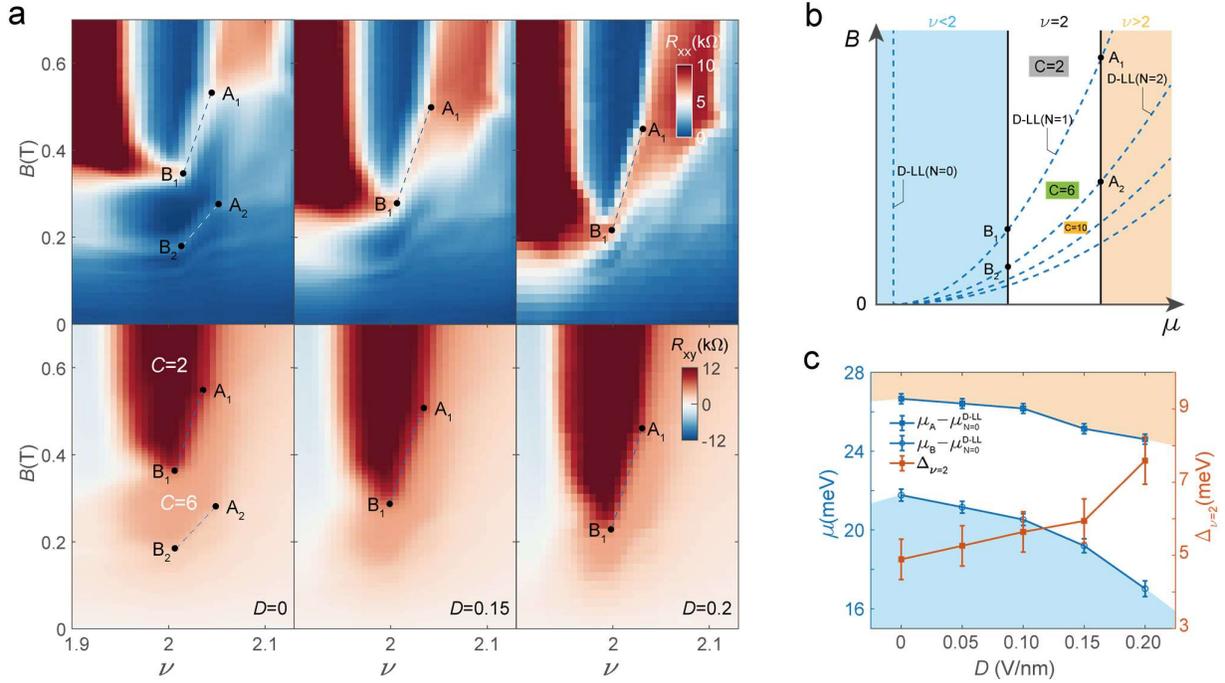

**Fig. 2 | Band crossings and extraction of the interaction driven gap at $\nu = 2$. a,** quantum oscillations at low magnetic field shown by longitudinal resistance $R_{xx}$ and transverse Hall resistance $R_{xy}$ at a variety of displacement field. Blue dash lines mark $R_{xx}$ peaks and $R_{xy}$ transitions for different Chern number $C = 4N + 2$ in magnetic field $B$, which corresponds to a phase where half-filled D-LLs cross the correlated gap of the flat band. **b,** schematics of band crossing between D-LLs and the moiré flat band. Horizontal and vertical axes denote chemical potential and perpendicular magnetic field. We ignore the negative compressibility of the flat band at $\nu < 2$. Black solid lines mark the band edges A and B of the surrounding correlated gap at $\nu = 2$. $A_1$ and $B_1$, $A_2$ and $B_2$ are points where 1st D-LL and 2nd D-LL cross with band edges A and B, respectively. **c,** extracted chemical potential of hte band edges relative to zeroth D-LL and correlated gap $\Delta(\nu = 2) = \mu_A - \mu_B$ as a function of displacement field. The extraction is carried out with D-LL index $N = 1$. Errors are from the uncertainty of magnetic field acquired in (a) for crossing points $A_1$ and $B_1$.

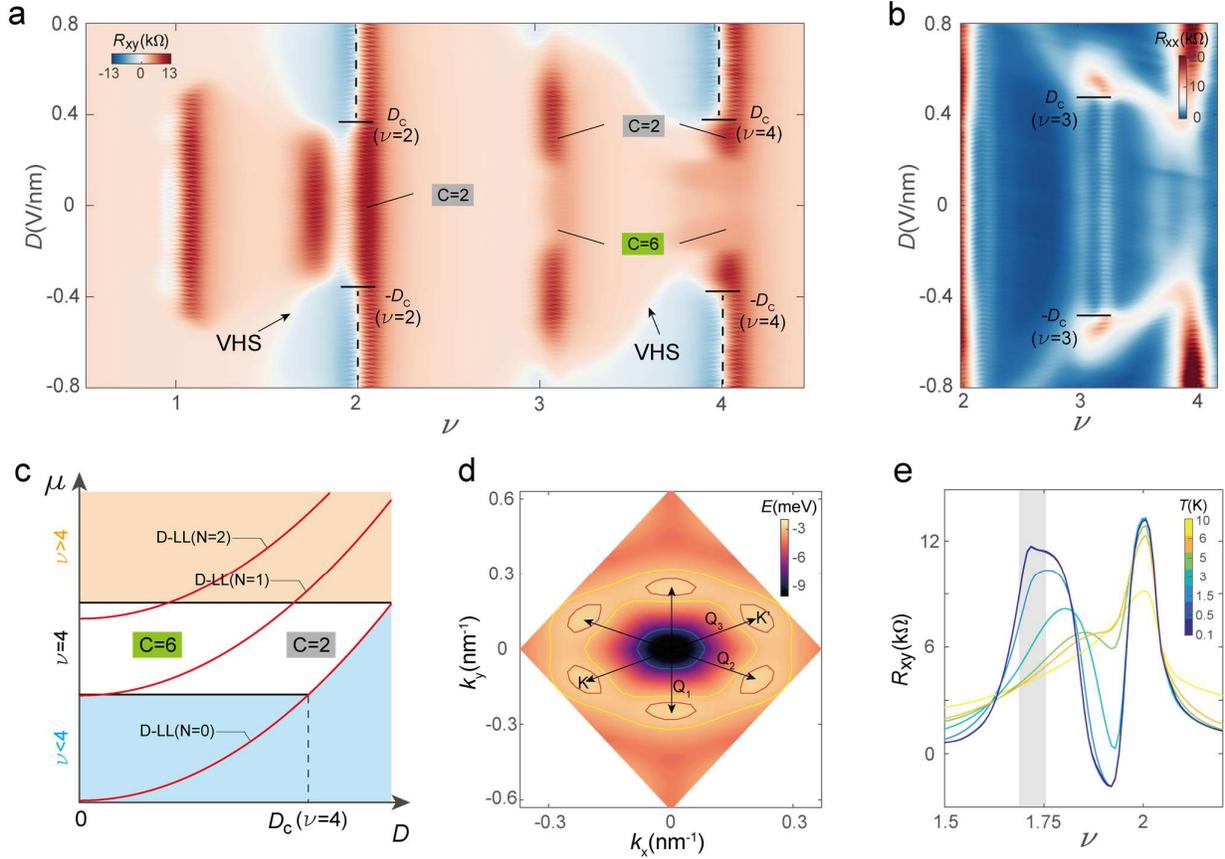

**Fig. 3 | Displacement field driven Dirac cone shifting and Fermi surface nesting. a,** color plot of transverse Hall resistance $R_{xy} = [R_{xy}(B) - R_{xy}(-B)]/2$ as a function of displacement field $D$ and charge filling $\nu$ at $|B| = 1T$. Vertical dash lines trace divergent $R_{xy}$ and cease at the as-marked horizontal solid lines which correspond to critical displacement field $D_c(\nu)$. **b,** color plot of longitudinal resistance $R_{xx}$ as a function of displacement field $D$ and charge filling $\nu$ at $B = 4T$ and $T = 2K$. At $\nu = 3$, $D_c$ is defined where the $R_{xx}$ dip vanishes. **c,** schematics of how the band crossings are affected by displacement field around $\nu = 4$. Horizontal and vertical axes denote displacement field and chemical potential, respectively, where zero energy is assigned to charge neutrality of moiré flat band. The flat band energy (denoted with blue shade) changes when the flat band at $|D| < |D_c(\nu = 4)|$ is ignorable compared with the Dirac cone shifting. We depict that the Dirac cone shifts quadratically in energy with respect to $D$. At $|D| > |D_c(\nu = 4)|$, flat band boundary is defined by fully hybridized Dirac cone. **d.** interacting band structure for valence band of $\nu = 2$ in moiré Brillouin zone through our Hartree-Fock simulation. The energy of $E = -3.154 meV$ corresponds to $\nu = 5/3$ filling, as marked by yellow contours. $\nu = 5/3$ is close to van Hove singularity where six hole pockets merge into annular Fermi sea with a formation of saddle points. There are three CDW wavevectors (Q1, Q2, Q3) connecting moiré K and K' points in a three-fold rotation symmetric way, as displayed in the contour plot. **e,** Hall resistance $R_{xy}$ versus charge filling at various temperature from $T = 0.1K$ to $T = 10K$ at $B = 0.5T$.

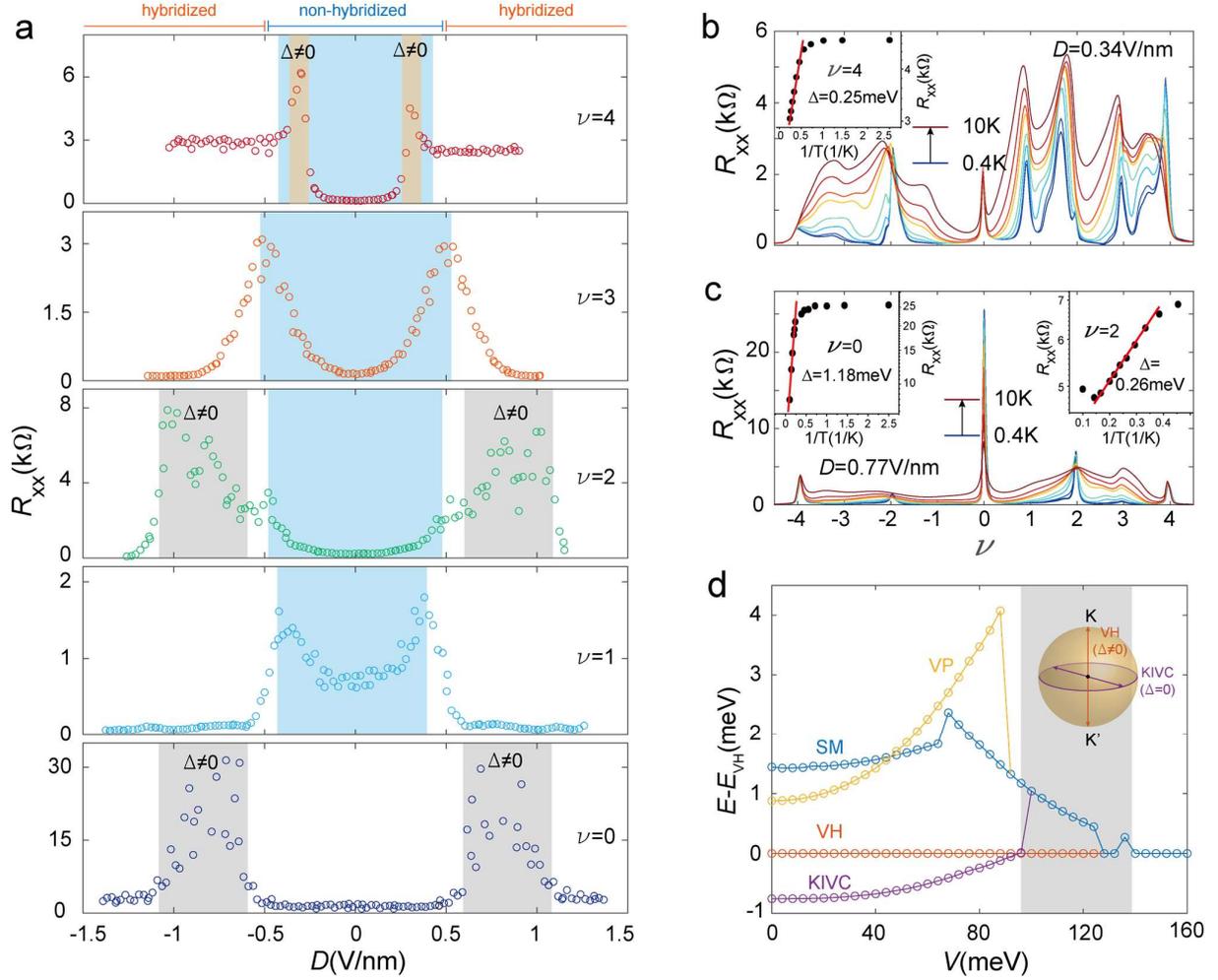

**Fig. 4 | Displacement field induced phase transitions. a**, four-probe resistance $R_{xx}$ as a function of displacement field $D$ at zero magnetic field. Blue shades indicate moderate displacement field regime where the Dirac cone and the flat band electrons are not hybridized. Beyond this regime, at higher $D$ Dirac cone and flat band are highly hybridized. The determination of hybridization regime is not available for $\nu = 0$ in our experiment. Gray and yellow shades denote gapped states in hybridized and non-hybridized regime, respectively. **b, c**, temperature-dependent behaviors of $R_{xx}$ at $D = 0.34 V/nm$ (b) and $D = 0.77 V/nm$ (c). The inset figures plot Arrhenius fitting $R \propto exp(\Delta/2kT)$, where $k$ is Boltzmann constant. **d**, energy difference for competing ground states of semimetal (SM), valley polarization (VP) and Kramer intervalley coherence (KIVC) as compared to valley Hall (VH) state. Horizontal axis is potential difference applied by displacement field. The inset shows Bloch sphere for VH and KIVC state, where VH is Ising antiferromagnet and KIVC is XY antiferromagnet. At $V < 100 meV$, KIVC state is with lowest energy; while at an interval of $100 meV < V < 140 meV$ depicted by gray shading, VH is favored. The critical potential difference $V = 100 meV$ for KIVC and VH transition corresponds to $D = 0.6 V/nm$ with dielectric constant $\varepsilon = 4.2$ and distance $d = 0.7 nm$ between top and bottom graphene layers. At $V > 140 meV$, SM becomes the ground state.

Supplementary Information

**Dirac cone spectroscopy of strongly correlated phases in twisted trilayer graphene**

Cheng Shen,[1] Patrick J. Ledwith,[2] Kenji Watanabe,[3] Takashi Taniguchi,[3] Eslam Khalaf,[2] Ashvin Vishwanath,[2] Dmitri K. Efetov[1]

1. ICFO - Institut de Ciencies Fotoniques, The Barcelona Institute of Science and Technology, Castelldefels, Barcelona, 08860, Spain

2. Department of Physics, Harvard University, Cambridge, MA 02138, USA

3. National Institute for Materials Science, 1-1 Namiki, Tsukuba, 305-0044, Japan

## A. Methods

Device fabrication.

The whole stacking is fabricated by the van der Waals assembly technique. Graphite, hBN and monolayer graphene flakes are mechanically exfoliated on $SiO_2$(285nm)/$Si^{++}$ substrates and chosen with optimal thickness by an optical contrast. A few nm thick graphite acting as top gate is first picked up at 100 °C by a propylene carbonate (PC) film which is placed on a polydimethyl siloxane (PDMS) stamp. All the other layers are subsequently assembled by repeating this procedure to make the final graphite/hBN/MATTG/hBN/graphite sandwich stacking as we showed in Fig.1a. For the assembling of MATTG, we precut the graphene flake into three separate pieces to reduce strain which is otherwise introduced during the picking up process. The second and third graphene layers are rotated with a target angle around 1.6° and -1.6°, respectively, to achieve the mirror symmetry. The whole stacking is finally released on a $SiO_2$(285nm)/$Si^{++}$ chip at 180°C and further patterned into a Hall bar geometry by following a typical ebeam lithography and $CHF_3/O_2$ reactive ion etching process. Dual gates and Hall bars are edge-contacted with Cr(5nm)/Au(50nm) metal.

Measurements.

The electronic transport measurements are carried out in a dilution fridge and at the base temperature of 35mK unless specified. A standard lock-in technique with excitation frequency $f$=17.111Hz and ac current <10nA is employed by using Stanford Research SR860. The ac current is applied through 10M ohm resistor. Local top and bottom gate are connected to source meters 2400 to tune carrier density and displacement field. We further apply global gate voltage (10V) to $Si/SiO_2$ to reduce the resistance of leads that are not gated by local graphite gates. The signals are taken after low-pass RC and LC filters which are used to reduce electron temperature.

Estimation of chemical potential, charge density and twist angle.

Based on the Fermi level location in $N$th D-LL and split flat bands, the phase diagram is composed of four types of phases (Extended Data Fig. 1): phase I: Fermi level $E_F$ located inside compressible $N$th D-LL and incompressible correlated flat band gap, corresponding to $N$th D-LLs crossing with correlated gap as we interpret in main text; phase II: $E_F$ located inside both compressible $N$th D-LL and correlated flat band, corresponding to $N$th D-LLs crossing with correlated flat band; phase III: $E_F$ located inside incompressible D-LL gap and compressible correlated flat band; phase IV: $E_F$ located inside both incompressible D-LL and correlated flat band gap. Phases I and II host $R_{xx}$ peak as a result of partially filling compressible D-LL. Phase IV hosts quantized $R_{xy}$ plateau. By changing magnetic field, phase IV with $R_{xy} = h/(4N - 2)e^2$ transitions to that with $R_{xy} = h/(4N + 2)e^2$. This transition corresponds to phase I.

Chemical potential of correlated flat band $\mu^f$ with respect to charge filling $\nu$ is estimated through phase II following a same method as in ref.[1] When $N$th D-LL crosses with flat band, the chemical potential of flat band $\mu^f$ relative to Dirac cone vertex equals to $\mu_N^{D-LL} - \mu_{N=0}^{D-LL}$ where $\mu_N^{D-LL} - \mu_{N=0}^{D-LL}$ is chemical potential difference between $N$th D-LL and zeroth D-LL that can be extracted experimentally via $\mu_N^{D-LL} - \mu_{N=0}^{D-LL} = v_F\sqrt{2e\hbar N sgn(N)B}$. Here, $v_F$ is Fermi velocity of Dirac cone, $sgn(N)$ is the sign of LL index $N = 0, \pm 1, \pm 2, \pm 3 \ldots$ and $B$ is perpendicular magnetic field. Charge density in Dirac cone when $N$th D-LL is half filled is obtained as $n^D = 4N\frac{B}{\phi_0}$ where $\phi_0 = \frac{h}{e}$ is the flux quantum and factor 4 denotes four-fold spin and valley degeneracy of D-LL. When D-LL is half filled, longitudinal resistance $R_{xx}$ reaches maximum. Hence, band crossing between half-filling D-LLs and flat band is easily recognized with pronounced $R_{xx}$ extrema through which the corresponding magnetic field $B$ can be captured.

We calculated the charge density $n^D \approx 0.5 \times 10^{11} cm^{-2}$ at integer filling $\nu = 2$ and $n^D \approx 1.2 \times 10^{11} cm^{-2}$ at $\nu = 3$ of flat band at zero displacement field. The exact filling of flat band is $\nu$ defined by relation $\nu = 4n^f/n_s^f$ ($n^f$ is the charge density of flat band and $n_s^f$ corresponds to full filling of flat band). At lower displacement field $D$ where flat band and Dirac cone coexist, $n^f$ is obtained as $n^f = n^T - n^D$ ($n^T$ is the total carrier density which is acquired through Hall density or gate capacitance). Alternatively, at each filling $\nu = \pm 2, 3, \pm 4, 5/3, 11/3$, trajectories of Hall plateau or longitudinal resistance $R_{xx}$ minimum (or maximum) with different slopes $\frac{dn}{dB} = (4N+2)e/h$ converge at zero magnetic field in a $n - B$ space. $n^T$ of the converging point is rightly $n^f$ for these integer or fractional fillings of flat band. At higher displacement field where flat band and Dirac cone are maximally hybridized, $n^f$ is directly obtained as $n^f = n^T$.

To extract the twist angle, we use the relation $n_s^f = 8\theta^2/\sqrt{3}a^2$, where $a = 0.246 nm$ is the lattice constant of graphene.

Extraction of correlated gap.

The precise measurement on correlated flat band gap at integer filling relies on precise determination of the crossing points $A_N$ and $B_N$ in $n - B$ phase diagram. In Extended Data Fig. 1, the phase boundary between phase III and IV, or I and II, corresponds to the state of $E_F$ touching split flat band edge, the former of which produces resistive $R_{xx}$ peaks in proximity to integer fillings due to D-LL edge state scattering with unfrozen flat band charges in the edges of split flat band (Fig. 2a), and the latter corresponds to $A_N$ and $B_N$. In the meanwhile, phase I and phase boundary between III and IV also converge on $A_N$ and $B_N$. At critical points $A_N$ and $B_N$, the transition of $R_{xy}$ plateaus with $C = 4N \pm 2$ in magnetic field, which is the characteristic of phase I, disappears. In addition, trajectory of resistive $R_{xx}$ peaks of phase I has a slope of $\frac{dn}{dB} = 4Ne/h$ in $n - B$ phase diagram, but that of phase boundary between III and IV inherits a slope of $\frac{dn}{dB} = (4N - 2)e/h$ from phase IV. This gives rise to a slope transition at $A_N$ and $B_N$. Notably, as flat band electrons exhibit negative compressibility in proximity to integer fillings, $B_N$ appears at the minimum of magnetic field in $n - B$ phase diagram. Such features help to precisely identify $A_N$ and $B_N$. Correlated gap is extracted through the as-acquired magnetic field at $A_N$ and $B_N$. The error of our measurement is determined by the $R_{xx}$ peak and $R_{xy}$ transition width in magnetic field of phase I, which are related to D-LL broadening that is associated with device disorders and magnetic field.

Note that we measure the gap with changed magnetic field. By varying D-LL index, the impacts of magnetic field change on correlated gap can be revealed. At $\nu = 2$ and $D = 0$, when D-LL index changes from $N = 1$ to $N = 2$, magnetic field correspondingly changes from around 0.45T to 0.25T (Fig. 2a). This induces gap decreasing by around 0.5meV, which is much smaller than the extracted correlated gap (see main text). The magnetic field change for correlated gap measurement is around the same order or even less (Fig. 2a). Note due to the Dirac cone shifting in displacement field, at elevated displacement field, phase I appears at a lower magnetic field and thereby the correlated gap would be measured with a bit smaller size. This leads to a slight underestimation of D-induced correlated gap enhancement at $\nu = 2$.

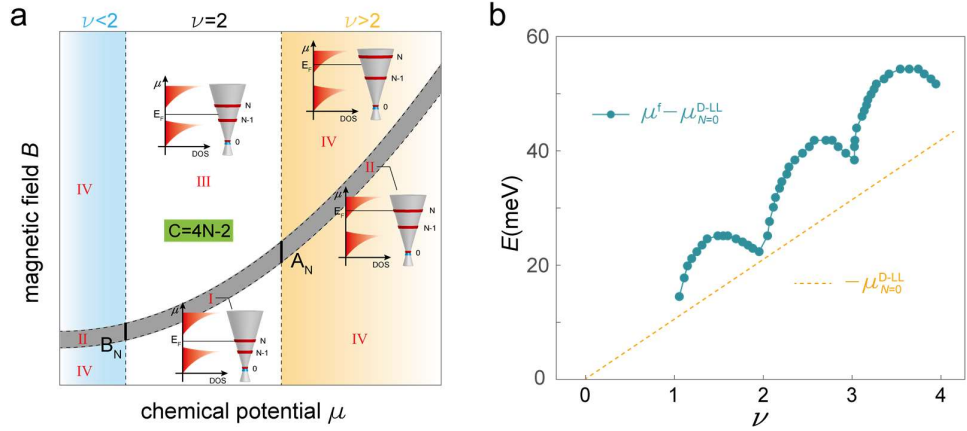

**Extended Data Fig. 1 | Phase diagram in perpendicular magnetic field and chemical potential measurement at $D = 0V/nm$. a**, four types of phases in perpendicular magnetic field when Dirac cone coexists with moiré flat band. The dash lines denote phase boundaries. Phases I and II correspond to partially filling $N$th D-LL, which shows finite band broadening due to disorder effects. The phase boundary between phase I and II is specially marked with solid line, pointing to critical points $A_N$ and $B_N$ as we discussed in main text. For each phase, there is a schematic illustration of filling status at Fermi level $E_F$ for flat band and D-LL. The emergence of correlated gap is shown to arise from flat band splitting. **b**, chemical potential measurement at $D = 0V/nm$. The dark cyan dots show energy difference between flat band and Dirac cone vertex, which is obtained via features of phase II with D-LL index $N = 1$. The yellow dash line denotes a tentative plotting of Dirac cone shifting as a function of charge filling, where the zero energy corresponds to energy of flat band charge neutrality.

## B. Correlated states at other fillings

We use our spectroscopy to reveal correlated states of other fillings at zero displacement field. For $\nu = -3$, at high magnetic field where only zeroth D-LL crosses with flat band, the magnitude of Hall resistance $R_{xy}$ is far below $h/2e^2$, indicative of a compressible metallic state (Extended Data Fig. 2). For $\nu = -2$, distinct from $\nu = 2$, $R_{xx}$ showing resistive peak rather than dip, reveals the presence of a few thermal-activated or unlocalized charges at our base temperature (Extended Data Fig. 2), indicating that a tiny or nodal gap is hosted by $\nu = -2$. This conforms to the observed particle-hole asymmetry in MATBG that $\nu = -2$ is usually with a much smaller resistance than $\nu = 2$.[2,3,4]

Fractional filling correlated state between $\nu = 3$ and $\nu = 4$ shows similar features to integer fillings $\nu = 2, 3, 4$ and fractional filling $\nu = 5/3$ with $R_{xx}$ minimum and nearly quantized $R_{xy} = h/[(4N+2)e^2]$ plateaus (Extended Data Fig. 2 and Extended Data Fig. 3). For high filling where carrier density in Dirac cone can't be ignored, filling $\nu$ deviates a lot from the real filling fraction of flat band. We identified $\nu^f = 11/3$ for the real fractional filling of flat band (Extended Data Fig. 3).

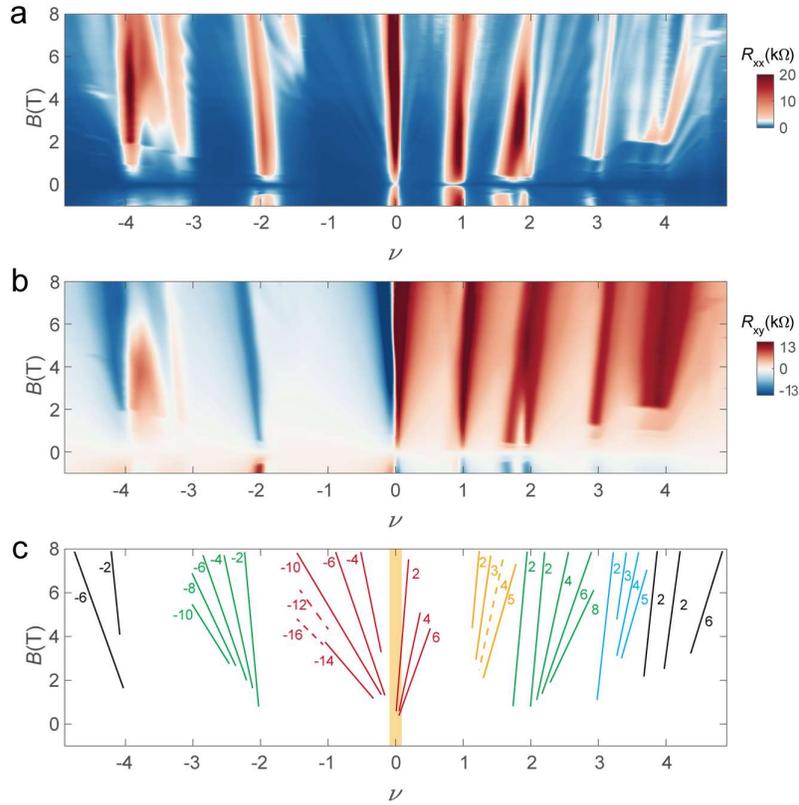

**Extended Data Fig. 2 | Landau fan diagram and Landau level crossing at $D = 0V/nm$. a, b,** Landau fan diagram shown by longitudinal resistance $R_{xx}$ and transverse Hall resistance $R_{xy}$. **c,** schematic of Landau level structure as observed in panel (a) and panel (b).

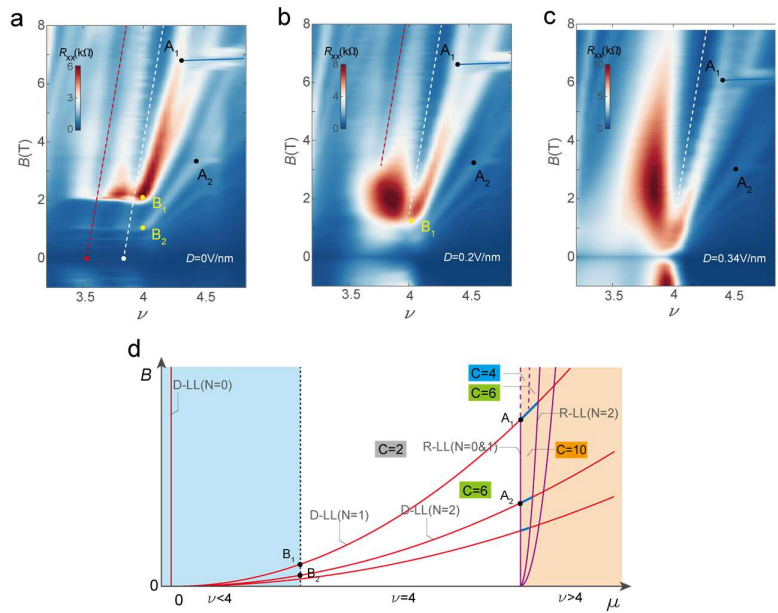

**Extended Data Fig. 3 | Correlated states at fractional filling 11/3 and band crossing around $\nu = 4$. a, b, c,** Landau fan diagram shown by longitudinal resistance $R_{xx}$ at various displacement field. The dash lines colored red and white trace $R_{xx}$ minimum with $C = 2$ hosted by fractional and full fillings of flat band, respectively. Solid blue lines denote gap closing for the remote band Landau level (R-LL). The ends of dash lines at zero magnetic field, denoted as solid dots, define a total filling $\nu$ corresponding to empty filling of Dirac cone, $\nu \approx$

3.53 for red dot, $\nu \approx 3.84$ for white dot. Thus, the real filling fraction of flat band $\nu^f$ for this fractional filling state is $\nu^f \approx 4 \times \frac{3.53}{3.84} \approx \frac{11}{3}$. The lower edge of remote band and upper edge of flat band are marked as point 'A' and 'B' with subscript '1' and '2' denoting the D-LL index. Panel (c) is taken at 1.8K. **d**, schematic of band crossing around $\nu = 4$. Red and violet lines represent D-LLs and R-LLs, respectively. Dash violet lines denote symmetry broken R-LLs. The blue lines (corresponding to blue lines in panel (a), (b) and (c)) as one part of the compressible D-LLs here, close the incompressible R-LL gaps.

### C. Unchanged $v_F$ in moderate displacement field

We present evidence of little changed $v_F$ of Dirac cone in moderate displacement field which is assumed in the chemical potential and gap extraction with nonzero moderate $D$. Firstly, in interacting simulations (supplementary section J), we didn't find any signature of reconstruction for Fermi velocity $v_F$ of Dirac cone when moderate displacement field is applied. Secondly, the critical displacement field $D_c(\nu = 2)$ acquired through $D$-induced $R_{xy}$ transitions at $\nu = 2$ is well consistent with the results of chemical potential which are measured with constant $v_F$ (Fig. 2c).

### D. Estimation of $\nu = 4$ single-particle band gap

We use the band crossing method at $\nu = 4$ as depicted for $\nu = 2$. Due to the high energy of remote band edge, the crossing happens at much higher magnetic field than $\nu = 2$, where Landau levels of remote bands (R-LLs) are formed. Crossing between D-LLs and R-LLs induces gap closing for the original incompressible Landau level gaps. Besides the features of band crossing we discussed in main text, tracking the LL gap closing can also help us to determine the exact crossing points A in $\nu - B$ diagram. Via $\mu_A - \mu_{N=0}^{D-LL} = v_F\sqrt{2e\hbar N sgn(N) B_A}$ and $\mu_B - \mu_{N=0}^{D-LL} = v_F\sqrt{2e\hbar N sgn(N) B_B}$, we extract at $D = 0$, $\mu_{A,\nu=4} - \mu_{N=0}^{D-LL} = 94.9 \pm 0.7 meV$, $\mu_{B,\nu=4} - \mu_{N=0}^{D-LL} = 51.6 \pm 1.8 meV$; at $D = 0.2V/nm$, $\mu_{A,\nu=4} - \mu_{N=0}^{D-LL} = 92.6 \pm 0.7 meV$, $\mu_{B,\nu=4} - \mu_{N=0}^{D-LL} = 45.7 \pm 2.4 meV$; at $D = 0.34V/nm$, $\mu_{A,\nu=4} - \mu_{N=0}^{D-LL} = 89.5 \pm 0.7 meV$, with constant $v_F = 10^6 m/s$. It is clear that both $\mu_{A,\nu=4}$ and $\mu_{B,\nu=4}$ decreases with $D$, signifying split Dirac cone shifting as we discussed in main text.

The most difference of chemical potential measurement at $\nu = 4$ is we need to consider the absent flat band broadening as a consequence of the absent Coulomb interaction energy $U'$ for remote band edge A since Fermi level doesn't reside in correlated flat band any more:

$$\mu_{A,\nu=4}(D) - \mu_{N=0}^{D-LL}(D) = E_0(D) + \Delta_{\nu=4}(D)$$

$$\mu_{B,\nu=4}(D) - \mu_{N=0}^{D-LL}(D) = E_0(D) + U'(D)$$

Here, $\Delta_{\nu=4}$ is single-particle band gap at $\nu = 4$, $E_0$ is single-particle energy difference from flat band edge to zeroth D-LL. Therefore,

$$\Delta_{\nu=4}(D) = \mu_{A,\nu=4}(D) - \mu_{B,\nu=4}(D) + U'(D)$$

$U'(D)$ is unclear in experiment. Instead, at critical field $D_c(\nu = 4)$, $E_0(D) = 0$ thereby the single-particle gap can be acquired as:

$$\Delta_{\nu=4}(D_c(\nu = 4)) = \mu_{A,\nu=4}(D_c(\nu = 4)) - \mu_{N=0}^{D-LL}(D_c(\nu = 4))$$

With D-LL index $N = 1$ and $D = 0.34V/nm$ which is approximate to $D_c(\nu = 4) \approx 0.37V/nm$, $\Delta_{\nu=4}$ can be obtained around

$$\Delta_{\nu=4}(D_c(\nu = 4)) \approx \mu_{A1,\nu=4}(D = 0.34V/nm) - \mu_{N=0}^{D-LL}(D = 0.34V/nm) = 89.5 meV$$

As $\mu_{A,\nu=4}(D_c(\nu = 4))$ is a bit smaller than $\mu_{A1,\nu=4}(D = 0.34V/nm) - \mu_{N=0}^{D-LL}(D = 0.34V/nm)$, $\Delta_{\nu=4}(D_c(\nu = 4))$ should be around but smaller than $89.5 meV$.

### E. Comparison on transport properties of two neighbor regions

We compare the transport properties of two neighboring regions. R1 is the region between Hall bar A and B where $R_{xx}$ in Fig. 1, Fig.3 and Fig. 4 are acquired, and Fig. 2 goes to R2 that is between B and C. Both of them have a same twist angle of 1.5° and show features of decreased $\mu_A - \mu_{N=0}^{D-LL}$, $\mu_B - \mu_{N=0}^{D-LL}$ and increased $\mu_A - \mu_B$ at $\nu = 2$ in displacement field. The difference is that R2 shows $R_{xx}$ minimum at fractional filling $\nu = 5/3$ rather than $R_{xx}$ maximum as in R1 (Extended Data Fig. 5), hosting a well-developed charge density wave state.

In Extended Data Fig. 4, we show four-probe resistance mapping plot in region R2. It shows a similar displacement field dependence as R1 (Fig. 1c and Fig. 4a).

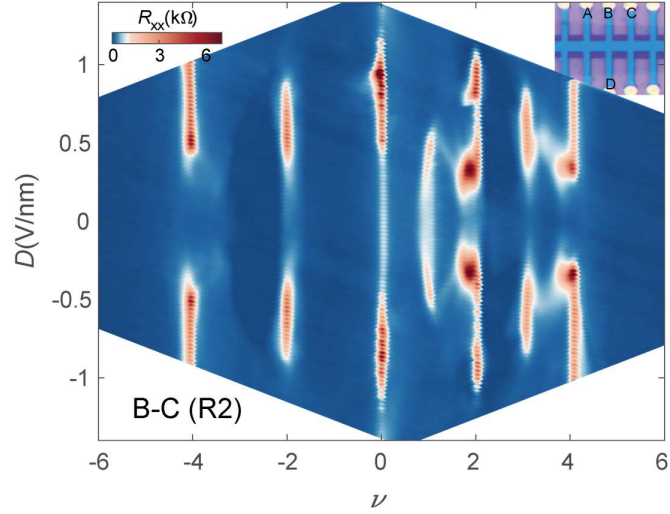

**Extended Data Fig. 4 | Additional $R_{xx} - \nu - D$ mapping plot from another part in our device.** Four-probe resistance $R_{xx}$ as functions of displacement field $D$ and charge filling $\nu$ acquired from Hall bars B-C at zero magnetic field. The inset optical microscopy picture shows measured Hall bars. $R_{xy}$ data throughout this paper are acquired from Hall bars B-D.

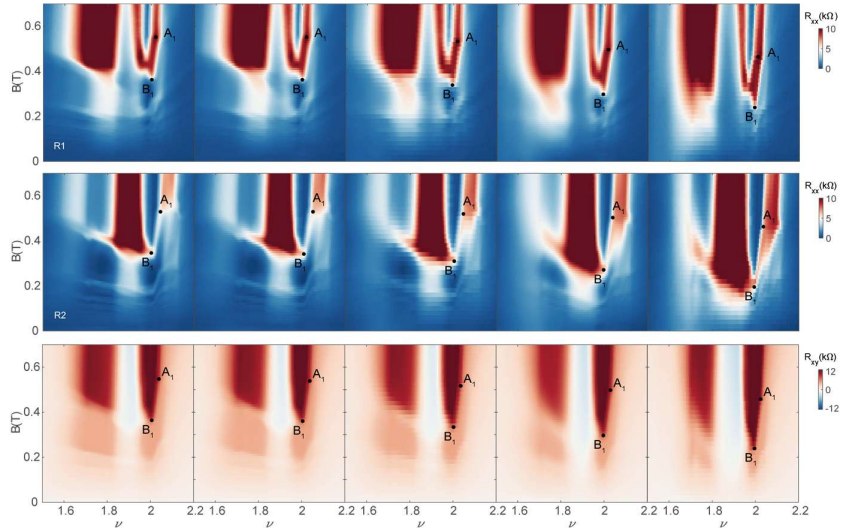

**Extended Data Fig. 5 | Zoom-in Landau fan diagram around $\nu = 2$ at a variety of moderate displacement field.** The top, middle and bottom panels are mapping plots of longitudinal resistance $R_{xx}$ of two neighbor regions R1 and R2 and Hall resistance $R_{xy}$, respectively. From the left to the right panel, displacement field $D$ is $D = 0, 0.05, 0.1, 0.15$ and $0.2 V/nm$ in sequence.

## F. Hierarchy of correlates states

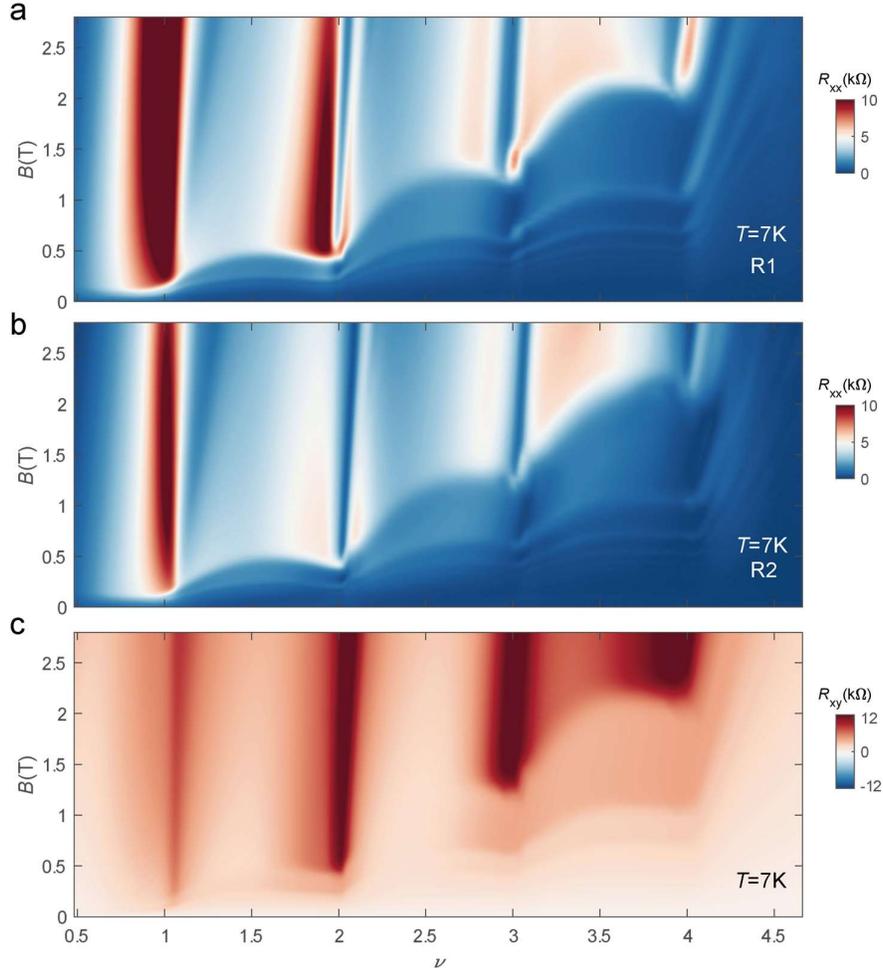

**Extended Data Fig. 6 | Hierarchy of correlates states. a-c**, longitudinal resistance $R_{xx}$ taken from neighbor region R1 (c), R2 (d) and Hall resistance $R_{xy}$ (e) versus perpendicular magnetic field $B$ and charge filling $\nu$ at $D = 0V/nm$ and $T = 7K$. $\nu = 2$ and $\nu = 3$ show robust incompressible gap states, i.e., feature $R_{xx}$ dip and $R_{xy}$ plateaus of $C = 4N + 2$. At $\nu = 1$, $R_{xx}$ peak is favored and $R_{xy}$ deviates from quantized value. There are no signatures of fractional filling $\nu = 5/3$ and $\nu = 11/3$, visualizing the hierarchy of correlated states for various fillings of flat band.

## G. Stronger electronic correlation at $D = 0.2V/nm$

At displacement field $D = 0.2V/nm$, most features of Landau fan resemble those at $D = 0V/nm$, i.e., being subject to coexisted non-interacting flat band and Dirac cone. Differences emerge at $\nu = -1$ and $\nu = 2$. As shown in Extended Data Fig. 7, Landau fan originating from $\nu = 2$ exhibits fully lifted degeneracy, distinct from the normal case in MATBG and $D = 0V/nm$ in MATTG where two-fold degeneracy with partial isospin symmetry breaking is favored. At $\nu = -1$, resistive state appears in high magnetic field. These features display much more pronounced spin-valley isospin symmetry breaking, indicative of stronger Coulomb interactions at $D = 0.2V/nm$ as compared to $D = 0V/nm$.

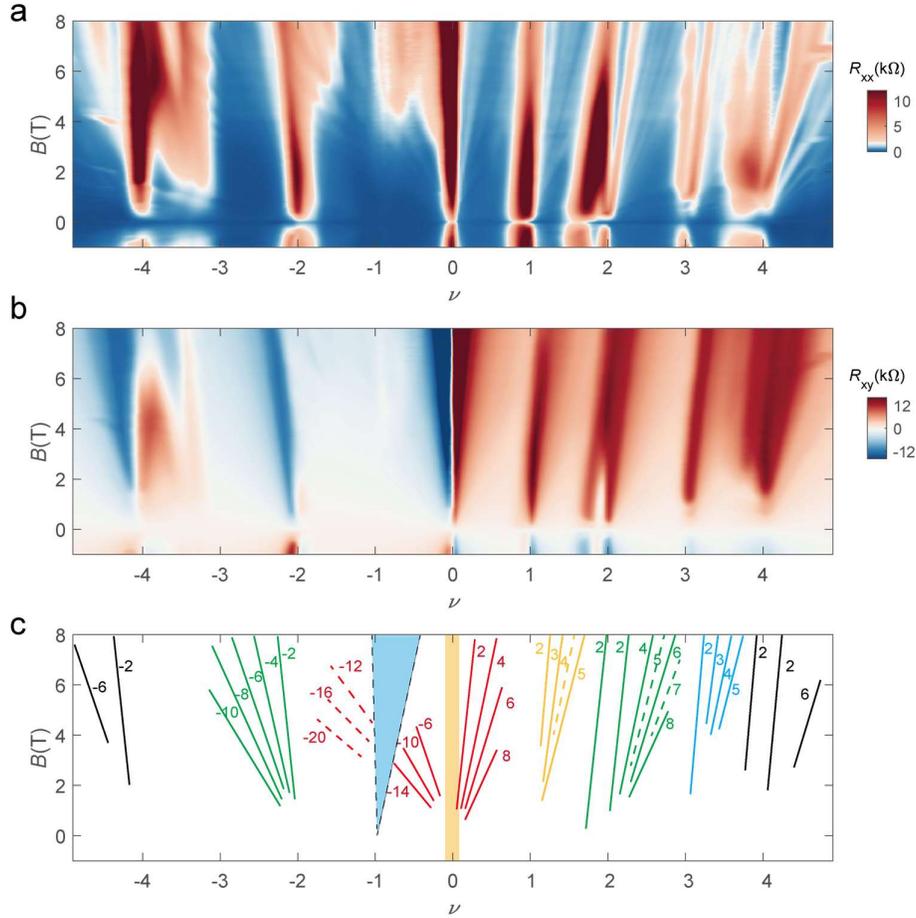

**Extended Data Fig. 7 | Landau fan diagram and Landau level crossing at $D = 0.2V/nm$. a**, **b**, Landau fan diagram shown by longitudinal resistance $R_{xx}$ and transverse Hall resistance $R_{xy}$. **c**, schematic of Landau level structure as observed in panel (a) and panel (b). The green dash lines denote odd Landau level filling factors from $\nu = 2$. Color shadings around $\nu = 0$ (yellow) and $\nu = -1$ (blue) represent resistive states in magnetic field.

## H. Extended temperature-dependent transport data

Extended Data Fig. 8 shows temperature-dependent $R_{xx}$ behavior at a larger temperature scale. At $D = 0.34V/nm$ close to $D_c \approx 0.4V/nm$, transport behavior of MATTG would be mainly dominated by flat band which means phenomena in MATBG will be reproduced in MATTG. For instance, in Extended Data Fig. 8 (a) and (b), Pomeranchuck effect is emergent at $\nu = -1$ with characteristics of vanished $R_{xx}$ peak and increased carrier density for phase boundary of isospin unpolarized states (IU) and symmetry-breaking isospin ferromagnets (IF) at low temperature.

At $D > D_c$, bandwidth $W$ of flat band is remarkably enhanced due to the strong coupling with shifted Dirac cone. The resulted largely reduced correlation strength disables isospin symmetry breaking at $1 < \nu < 2$, which is reflected by the crossed Landau fan from charge neutrality and disappeared resistance peak at $\nu = 1$ for $D = 0.77V/nm$ (Extended Data Fig. 9). Such features of $\nu = 1$ mimic $\nu = -1$ in MATBG.[4] We found $\nu = 1$ at $D = 0.77V/nm$ inherits features of Pomeranchuck effect such as vanished resistance peak at low temperature. The Pomeranchuck state being pronounced at $\nu = 1$ as a result of less strong electronic correlation, conforms to the hierarchy of correlated states for integer fillings of flat band.

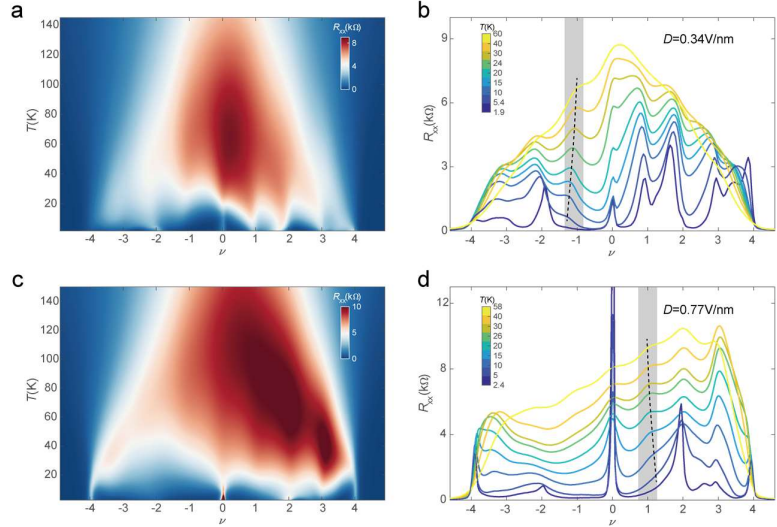

**Extended Data Fig. 8| Temperature dependence of four-probe resistance. a**, mapping plot of longitudinal resistance $R_{xx}$ as a function of charge filling $\nu$ and temperature $T$ at a nonzero displacement field $D = 0.34V/nm$. **b**, linecuts of $R_{xx}$ versus $\nu$ taken from panel (a). **c, d**, mapping plot and linecuts of $R_{xx}$ at $D = 0.77V/nm$. The gray shadings denote vanishing and shifted $R_{xx}$ peak at low temperature as a signature of Pomeranchuck effect.

## I. Overall Landau fan diagram at $D = 0.77V/nm$

At $D = 0.77V/nm$, Dirac cone is fully hybridized and its vertex is split into two, each of which is pushed to the edge of moiré flat band. As a result, Landau level crossing appears only for $|\nu| > 4$ as shown in Extended Data Fig. 9 that Landau fans originating from $|\nu| = 4$ show $C = 4N$ towards charge neutrality whereas $C = 4N + 2$ outwards. Another notable feature of Extended Data Fig. 9 is the response of $|\nu| = 2$ in magnetic field. At $B < 4T$, $|\nu| = 2$ shows Chern number $C = 0$; whereas at $B > 4T$, Chern insulators with $C = 0$ are absent instead replaced by one with $|C| = 2$.

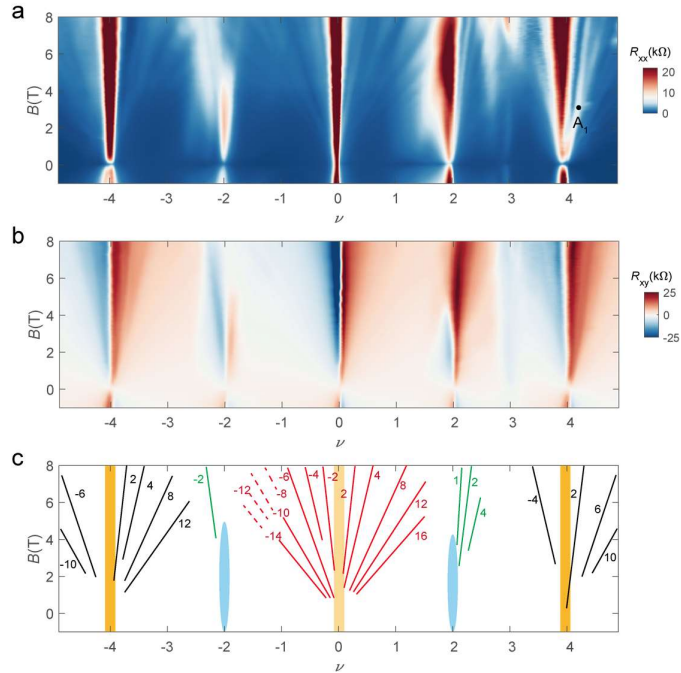

**Extended Data Fig. 9| Landau fan diagram and Landau level crossing at $D = 0.77V/nm$. a, b,** Landau fan diagram shown by longitudinal resistance $R_{xx}$ and transverse Hall resistance $R_{xy}$. **c,** schematic of Landau level structure as observed in panel (a) and panel (b). Landau levels fanning outward from $\nu = \pm 4$ show filling factor sequences of $\nu_{LL} = 4N + 2$, indicating Landau level crossing between Dirac cone and remote dispersive bands, while Landau levels fanning inward from $\nu = \pm 4$ show $\nu_{LL} = 4N$. Resistive states at $\nu = 0$, $\nu = \pm 2$ and $\nu = \pm 4$ are denoted with shadings colored yellow, blue and orange, respectively.

## J. Hartree-Fock simulations for MATTG

Twisted trilayer graphene (TTG) at zero displacement field consists of a twisted bilayer graphene subsystem (TBG) as well as a decoupled graphene subsystem. A large body of literature,[5] including analytical[6–9] and numerical[10–17] works, have argued that strainless magic angle TBG is well-described by generalized quantum Hall ferromagnetism in terms of a Chern-band basis. There are eight Chern bands, four of each sign $C = \gamma_z = \pm 1$. Within each Chern sector, the four bands are labeled by their valley $\eta_z = \pm 1$ and spin $s_z = \pm 1$. The quantum Hall ferromagnetic states correspond to fully filling some of these bands, though not necessarily bands with definite valley or spin quantum number – we may take linear combinations within each Chern sector.

At small nonzero displacement fields, these quantum Hall ferromagnets become coupled to the graphene Dirac cones leading to many of the effects described in the main text. Our discussion will summarize results from the detailed perturbation theory carried out in ref.[18] (see also ref.[19] which also found a transition from gapless KIVC to gapped VH order). We first focus on spinless TTG at charge neutrality for simplicity. Then we upgrade to $\nu = 2$ TTG which may be thought of as spinless charge neutral TTG on top of a spin polarized background (with an additional Hartree dispersion term).

At zero displacement field, the effective TBG Hamiltonian at charge neutrality is $h_{TBG}(\mathbf{k}) = h_x(\mathbf{k})\hat{\gamma}_x + h_y(\mathbf{k})\hat{\gamma}_y + \widehat{\Delta}(\mathbf{k})$. Here, $\gamma_{x,y}$ are Pauli matrices that flip the Chern number $C = \gamma_z = \pm 1$ but leave the valley $\eta_z = \pm 1$ invariant. The term $\widehat{\Delta}$ is a mean field potential coming the generalized ferromagnetic ordering of the TBG Chern bands; it is proportional to the interaction strength. The two states that we will focus on are the Kramers' intervalley coherent (KIVC) state which is off diagonal in valley, $\widehat{\Delta} \propto \gamma_z \eta_{x,y}$, and the valley Hall (VH) state which has $\widehat{\Delta} \propto \gamma_z \eta_z$. Both of these states have $\widehat{\Delta} \propto \gamma_z$. As a result the mean-field pairing potential anticommutes with the rest of the TBG dispersion for these states; this favors them energetically.[6,7] The charge gap may then be computed as the minimum of

$$2\sqrt{h_x^2(\mathbf{k}) + h_y^2(\mathbf{k}) + \Delta_0^2(\mathbf{k})}$$

over the Brillouin zone, where $\Delta_0$ is the magnitude of $\widehat{\Delta}$.

The zero-field graphene Hamiltonian is

$$h_{GRA}(\mathbf{k}) = h_{GRA}^+(\mathbf{k})\hat{\eta}_+ + h_{GRA}^-(\mathbf{k})\hat{\eta}_-, \quad h^\pm(\mathbf{k}) = \pm(\mathbf{k} - \mathbf{K}_\pm) \cdot \hat{\boldsymbol{\gamma}},$$

where $\mathbf{K}_{+(-)}$ is the momentum moiré K (K') point. Crucially, the Dirac points in different valleys have different moiré crystal momentum. As a result, they cannot develop intervalley coherent order.

The displacement field $V$ induces a tunneling between the two subsystems. Here, we write $V$ to signify the potential difference between the top and bottom layers of MATTG after screening – this is the natural input to theoretical calculations. The relationship between $V$ and the bare, experimental, displacement field $D$ in principle may be obtained through a full self-consistent calculation but here we simply choose an empirical value based on previous studies of moiré systems. We take the conversion factor $V = (0.166 \text{ nm}) eD$ where $e > 0$ is the magnitude of the electron charge. This is motivated by a comparison between experimental and Hartree Fock studies of twisted monolayer-bilayer graphene where a fractional filling Chern insulator appears for a narrow range of displacement field.[20] Note that ref.[21] used $V = (0.1 \text{ nm})eD$ in the context of twisted double bilayer graphene though this system has four layers.

Turning on a displacement field immediately results in a reconstruction of the non-interacting band structure (see Fig. 1b). The TBG and graphene Dirac cones fully hybridize and lift off of zero energy at the K point, though the system remains gapless due to a Dirac point at the K' point. We emphasize that this effect of the displacement field relies crucially on the lack of a charge gap at the K point in the twisted bilayer graphene subsystem. When the TBG sector has a charge gap, small displacement fields only lead to a small mixing between the TBG and graphene subsystems instead of an immediate, maximal, hybridization. See Extended Data Fig. 1 for Hartree Fock band structures of MATTG with and without a displacement field.

We may understand the effect of the displacement field on states with a TBG charge gap through perturbation theory; we will see that it can be understood in a similar way to the 'superexchange' mechanism in cuprates. The displacement field tunnels TBG electrons to the graphene sector and vice versa. The tunneling is Pauli blocked if the occupations of the subsystems are the same. However, if the occupations are not the same the system may lower its energy by slightly mixing the subsystems. To favor the latter scenario, the displacement field induces a dispersion in both sectors that tries to make the occupations of the two subsystems less similar. In the cuprates, this leads to antiferromagnetism. For us, it is responsible for the increase of the flat band charge gap as well as the transition from the KIVC to a fully gapped VH state at nonzero displacement field as we now describe.

The TBG electrons gain a dispersion $h_V \propto V^2$ which has the opposite sign of the graphene dispersion so that it favors opposite occupation in the two subsystems. This dispersion combines with native TBG dispersion $h_x(\mathbf{k})\hat{\gamma}_x + h_y(\mathbf{k})\hat{\gamma}_y$ and may increase it or lower it depending on the twist angle. At large twist angles, the TBG dispersion resembles the graphene dispersion, but as the twist angle passes through the magic angle the dispersion changes sign. The angle $\theta = 1.50°$ is expected to be smaller than the angle at which the dispersion changes sign[18] and so we expect the displacement field to effectively increase the native TBG dispersion as we claimed above and in the main text. The increased dispersion will lead to the observed increase in band gap, but it will also increase the "Chern superexchange coupling" $J \propto h^2$ that favors states with $\hat{\Delta} \propto \gamma_z$. This superexchange coupling is also believed to be responsible for pairing skyrmions, setting the skyrmion effective mass, and setting the critical temperature for the skyrmion superconductor $T_c \propto J$.[7,22,23] The observed increase in $T_c$ with displacement field is consistent with the skyrmion mechanism for superconductivity and increased effective TBG dispersion with displacement field.

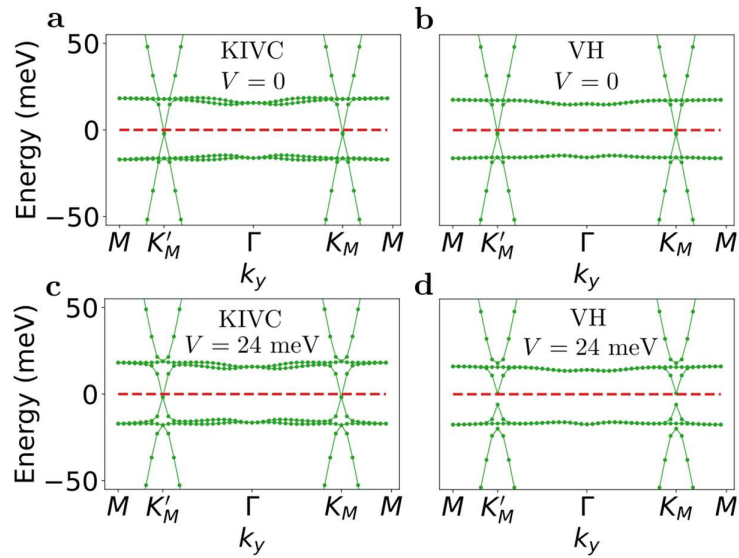

**Extended Data Fig. 10 | Self consistent Hartree Fock Band structures for spinless MATTG at charge neutrality**. We include band structures for the Kramers' intervalley coherent (KIVC, panels (a) and (c)) and valley Hall (VH, panels (b) and (d)) states at $V = 0$ and $V = 24\ meV$ respectively. Note that the inclusion of both graphene valleys results in graphene Dirac cones at both the $K_M$ and $K'_M$ points. There is no immediate full TBG-graphene hybridization at the K points unlike non-interacting MATTG due to the charge gap in the TBG sector. For the valley Hall state, the graphene Dirac cone gains a mass $m \propto V^2$.

Similarly, the graphene electrons gain a dispersion that is opposite to the TBG dispersion. The TBG dispersion is dominated by the mean-field potential. The induced graphene dispersion was computed in ref.[18] and at the K point is given by

$$m_G = -\frac{V^2|t|^2}{4U}\frac{\hat{\Delta}}{\Delta_0}.$$

Here, $U \approx 18$ meV comes from the TTG interaction strength and $|t| \approx 0.51$ is a tunneling matrix element that measures the spatial overlap between the TBG and graphene wavefunctions. Recall $\hat{\Delta} \propto \gamma_z$ for the KIVC and VH states and therefore $m_G$ anticommutes with the graphene dispersion; for these states $m_G$ functions as a Dirac mass that can gap out the graphene subsystem. The KIVC mass can be ignored since as discussed above the graphene Dirac cones are not at the same momentum in each valley. The VH mass term is effective; it generates a gap for the graphene Dirac cones. Because the VH mass term can be energetically satisfied, the energy of the VH state lowers relative to the KIVC state by an amount $\propto m_G^2$. This results in a transition from a KIVC state, where the graphene electrons are semimetallic, to a VH state where the graphene electrons are gapped.[18,19] The transition is dependent on the ratio between intra-sublattice tunneling and inter-sublattice tunneling κ, which is largely unknown. Ref.[18] estimates κ = 0.58 and a critical displacement field of $V = 100$ meV or $D \approx 0.6$ V/nm.

We now comment on generalizing these results to $\nu = 2$. At $\nu = 2$, we may describe candidate ferromagnets with total Chern number zero by polarizing the same two flavors in each Chern sector and then half filling the remaining four bands. The flavor polarization is expected to correspond to a spin polarization, as valley polarization breaks time reversal symmetry which seems unlikely due to the proximate superconductors. The spin polarization may be aligned or anti-aligned in the two valleys depending on the sign of the intervalley Hunds coupling[22] but this does not matter for our purposes and we assume aligned spins in both valleys; one may always rotate the spin in the other valley to translate our conclusions to the spin-valley locked case. With this assumption, one spin is completely polarized, let's call this spin "down". The unpolarized, up, spin sector resembles charge neutral twisted bilayer graphene, though with the additional background charge density from the down spins.

The background charge induces a mostly flavor diagonal Hartree dispersion $h_0(k)$ for the TBG electrons which has a prominent dip at the Γ point (see ref.[24] for a detailed discussion). Because the Hartree dispersion is largely flavor diagonal, it simply adds onto the previously discussed spinless charge neutral TBG dispersion $\pm\sqrt{h_x^2(\boldsymbol{k}) + h_y^2(\boldsymbol{k}) + \Delta_0^2(\boldsymbol{k})}$. Near the magic angle, the Hartree dip causes the conduction band to have a minimum at the Γ point. The valence band maximum is very parameter dependent because the charge neutral TBG dispersion and the Hartree dip usually nearly cancel which results in a very flat band. The maximum may be at the Γ point but even if it is not the valence band is flat enough that the energy at the Γ point is a good approximation. Thus, up to a constant determined by the Hartree dip, we approximate the charge gap for $\nu = 2$ with the charge neutrality band gap at the Γ point of spinless TTG. Accordingly, we performed Hartree-Fock simulations for charge neutral spinless TTG and extracted the change in the band gap at the Γ point as a function of $V$; this is plotted in Extended Data Fig. 11. It is worth noting that, within Hartree Fock theory and with perfect spin polarization, only spin up electrons acquire the mass $m_G$ in the flat band VH state; the spin down electrons remain semimetallic. However, to our knowledge there is no symmetry that protects the gaplessness of the spin down electrons because $C_2T$ has been broken by the VH order. Thus there may be beyond-Hartree-Fock effects that lead to a small gap amongst the spin down electrons. We leave a microscopic calculation of the resulting spin down gap to future work. Note that the activated gap and resistance peak for $\nu = 2$ is much smaller than that of $\nu = 0$ (see Fig. 4a).

At zero displacement field, the background charge density from the polarized down spin simply shifts all of the graphene Dirac cones up in energy. However, by the same argument given above for the graphene mass, when a nonzero displacement field is turned on the system wants the down spin graphene electrons to be unoccupied so that displacement-field induced tunneling is not Pauli-blocked. This effect acts to shift the down spin graphene Dirac cones further up in energy, relative to zero displacement field. The additional shift is given by

$$\mu_\downarrow = \frac{|t|^2 V^2}{4U}.$$

Note that the experimental gap extraction assumes that the Dirac cones are not spin split. The above splitting can therefore lead to an underestimate of the gap increase with displacement field (see Extended Data Fig. 11). It also leads to the observed decrease in $\mu_B - \mu_{N=0}^{D-LL}$ (Fig. 2c).

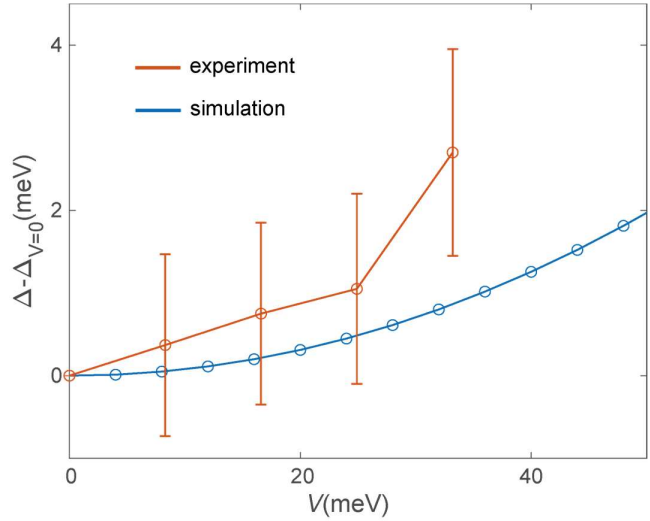

**Extended Data Fig. 11 | Gap enhancement at $\nu = 2$ with respect to displacement field $V$.** The blue line denotes gap enhancement for $\nu = 2$ approximated with charge neutrality band gap at the $\Gamma$ point of spinless TTG which is calculated with Hartree-Fock simulations. The orange line presents results extracted experimentally. Here, we use $V = (0.166 \text{ nm}) eD$ to convert displacement field $D$ to potential difference $V$ between top and bottom layers of TTG.

We now comment on the observed fractional filling state at $\nu = 5/3$. Our approach is to consider a Hartree Fock band structure at $\nu = 2$ and look for van Hove singularities upon hole-doping the valence bands. For simplicity, and motivated by an ignorance of other small scales that may change the order of the valence bands, we neglect the bare dispersion which enables us to do Hartree Fock without iterating for self-consistency. In this limit there are two valence bands that are each three-fold degeneracy. Because the valence bands are quite flat, we assume that the interaction will spontaneously choose one valence band to hole dope by 1/3 and thus ignore the degeneracy. Remarkably, we find that the top valence band at hole filling 1/3 is extremely close to a van Hove singularity where six hole pockets surrounding the $\Gamma$ point merge into a single annular fermi sea. Furthermore, the annular Fermi sea is very close to being nested with respect to the six wavevectors $(R_{2\pi/6})^m Q_1$. Here, $m = 0,1,2,3,4,5$; $Q_1 = k_\theta(0,-1)$ is a vector connecting the two moiré K points; and $R_{2\pi/6}$ is a rotation by $2\pi/6$. These vectors are precisely the translation breaking wavevectors that correspond to a $\sqrt{3} \times \sqrt{3}$ tripling of the unit cell. We therefore conclude that one third hole filling the top valence band – and spontaneously lifting the approximate three-fold degeneracy – puts the system in a state that is highly unstable to tripling the unit cell. This matches experimental observations of an insulating state at this filling that is related to a van Hove singularity.